\documentclass[twocolumn,aps,prl,bibnotes10pt,superscriptaddress]{revtex4}

\usepackage{amsmath}
\usepackage{graphicx}
\usepackage{ulem}
\usepackage{dcolumn}
\usepackage{epsfig}
\usepackage{bm}
\usepackage{array}
\usepackage[frenchb]{babel}
\usepackage{hyperref}
\hypersetup{
colorlinks=true,
citecolor=black,
linkcolor=red,
urlcolor=black
 , bookmarks=true,
 pdfmenubar=true
}

\usepackage{hyperref}
\usepackage{graphicx}
\usepackage{amsmath}
\usepackage{amsfonts}
\usepackage{amssymb}
\usepackage{xcolor}
\usepackage{bbm}
\usepackage{calrsfs}
\usepackage{dutchcal}
\usepackage{amsthm}

\newcommand{\new}[1]{\textcolor{blue}{#1}}

\newcommand{\ket}[1]{\ensuremath{|#1\rangle}}

\newcommand{\mean}[1]{\ensuremath{\big\langle #1 \big\rangle}}

\newcommand{\vect}[1]{\bm{#1}}

\newcommand{\be}{\begin{equation}}
\newcommand{\ee}{\end{equation}}

\newcommand{\beq}{\begin{eqnarray}}
\newcommand{\eeq}{\end{eqnarray}}

\newcommand{\J}{\hat{J}}

\begin{document}

\title{Heisenberg-limited noisy atomic clock \\ using a hybrid coherent and squeezed states protocol}
\author{Luca Pezz\`e and  Augusto Smerzi} 
\affiliation{QSTAR, INO-CNR and LENS, Largo Enrico Fermi 2, 50125 Firenze, Italy} 

\begin{abstract}
We propose a hybrid quantum-classical atomic clock protocol where the interrogation 
of an ensemble of uncorrelated atoms in a spin-coherent state
is used to feedback one (or more) spin-squeezed atomic ensembles toward their optimal phase sensitivity point. 
This protocol overcomes the stability of a single Ramsey clock and
it reaches a Heisenberg-limited stability while avoiding non-destructive measurements.
Analytical predictions are compared with numerical simulations of clocks operations including correlated $1/f$ local oscillator noise.
\end{abstract}

\maketitle
\date{\today}

Atomic clocks provide stable and accurate frequency and time references~\cite{LudlowRMP2015, PoliNC2013}.  
These are crucial for technological applications and fundamental research,
from relativistic geodesy~\cite{LisdatNATCOMM2016, McGrewNATURE2018, MehlstaublerRPP2018, KolkowitzPRD2016}
to the search for variations of the fine-structure constant~\cite{DereviankoNATPHYS2014, SannerNATURE2019, RobertsARXIV2019}.
Intense efforts are currently focusing 
on new strategies to further increase the stability of atomic clocks. 
These include the realization of low-decoherence lasers~\cite{ColeNATPHOT2013, NemitzNATPHOOT2016}, 
the decrease of interrogation dead times~\cite{TakamotoNATPHOT2011, SchioppoNATPHOT2016, ChouPRL2011}, 
and the reduction of phase estimation uncertainties below the standard quantum limit (SQL)~\cite{PezzeRMP2018}, 
$\Delta \theta_{\rm SQL} = 1/\sqrt{N}$, a bound
imposed by quantum mechanics when using $N$ classically correlated atoms~\cite{ItanoPRA1993, GiovannettiPRL2006, PezzePRL2009}.
In principle, the clock sensitivity can arbitrarily increase with the number of atoms. 
In realistic scenarios, however, $N$ is limited either by the use of specific non-scalable platforms 
or by decoherence effects, like collisional shifts and three-body recombinations.
The possibility to overcomes the SQL by creating tailored quantum correlations among the atoms 
is therefore attracting an increasing interest~\cite{PezzeRMP2018}. 
Experiments have explored the creation of spin-squeezed 
states~\cite{WinelandPRA1992, KitagawaPRA1993, WinelandPRA1994, MaPHYSREP2011, MeiserNJP2008, GilPRL2014}
with Bose-Einstein condensates~\cite{GrossNATURE2010, RiedelNATURE2010, BerradaNATCOMM2013}, trapped ions~\cite{BohnetSCIENCE2016} and
cold atoms~\cite{AppelPNAS2009, Schleier-SmithPRL2010, BohnetNATCOMM2014, HostenNATURE2016}, demonstrating proof-of-principle atom interferometers 
with sensitivities overcoming the SQL~\cite{PezzeRMP2018, Louchet-ChauvetNJP2010, LerouxPRL2010, HostenNATURE2016, BravermanPRL2019}. 
Squeezed states have reduced fluctuations of some chosen observable with respect to spin-coherent states 
and are far more robust to decoherence than GHZ or NOON states. Squeezed states are therefore good
candidates for magnetometers~\cite{SewellPRL2012, MuesselPRL2014, OckeloenPRL2013} and atom interferometers~\cite{SalviPRL2018} with performances overcoming the current technology. 
However, the possible advantages offered by squeezing in noisy atomic clocks, where  
the main source of noise are the fluctuations of the local oscillator (LO) during a Ramsey interrogation,
is still debated~\cite{AndrePRL2004, ShigaNJP2012, BorregaardPRL2013, KesslerPRL2014, MullanPRA2014, ChabudaNJP2016, SchulteARXIV2019}. 
The reason is that squeezed states allow to decrease the phase uncertainty $\Delta \theta$ below the SQL only when the unknown phase $\theta$
is sufficiently close to the optimal point $\theta_{\rm opt}$
of largest slopes of the Ramsey interference fringes, and rapidly degrade once $\theta$ drifts away.  
In addition, the more the state is squeezed, the narrower is the range of phase values where $\Delta \theta < \Delta \theta_{\rm SQL}$. 
These effects dramatically limit the usefulness of squeezed states to increase the long-term stability of noisy clocks, providing sensitivities
$\Delta \theta / \Delta \theta_{\rm SQL} \sim N^{1/6}$~\cite{AndrePRL2004},
which are far from the optimal Heisenberg limit $\Delta \theta_{\rm HL} / \Delta \theta_{\rm SQL} = 1/N^{1/2}$. The possibility to 
overcome this limitation has triggered many efforts
to develop protocols combining squeezing and non-destructive measurements~\cite{ShigaNJP2012, BorregaardPRL2013}.
Unfortunately, back-action effects introduce a loss of atomic coherence~\cite{KohlhaasPRX2015, ColangeloNATURE2016, HammererRMP2010} that limits the performances of these protocols.

\begin{figure} [b!]
\centering
\includegraphics[width=1\columnwidth]{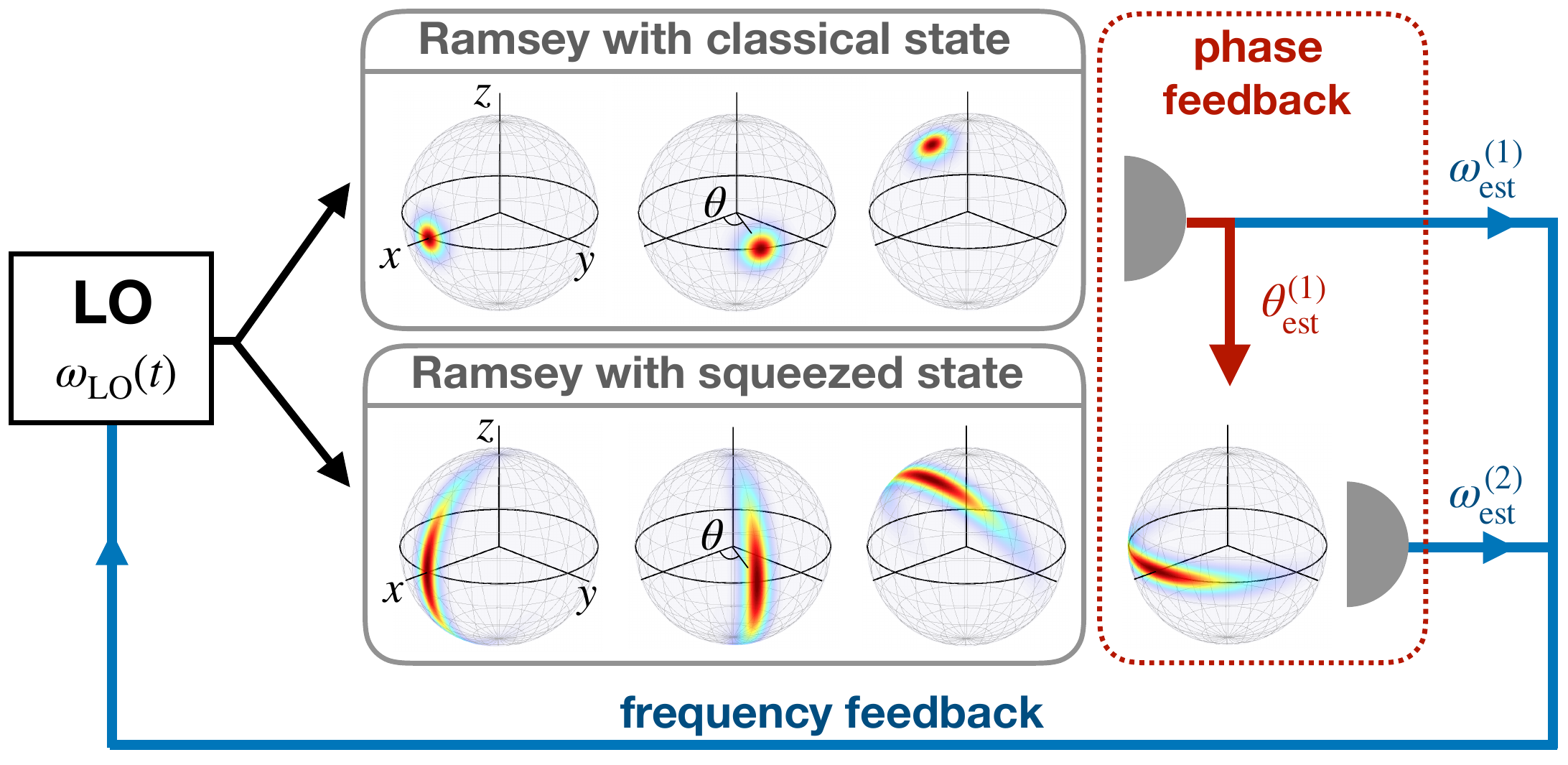}
\caption{Hybrid clock with two atomic ensembles interrogating the same local oscillator.
Both Ramsey interferometers consist of state preparation (left-hand side showing the Husimi distribution), phase shift (central - the accumulated phase is the same for both ensembles)  
and $\pi/2$ rotation about the $x$ axis (right).
Ramsey 1 uses a coherent spin state, while Ramsey 2 uses a squeezed state.
The readout of Ramsey 1 is used to phase-feedback (red line) Ramsey 2 in order to bring the squeezed state to its most sensitive phase estimation
optimal region with a final rotation about the $y$ axis
The frequency estimations are combined to steer the frequency of the local oscillator (blue lines).}
\label{fig1}
\end{figure}

Here we propose an atomic clock protocol where a single coherent and a few 
squeezed states, all created in different atomic ensembles, are interrogated by the same LO -- 
Fig.~\ref{fig1} illustrates the case of a two ensembles scheme.
The central idea is to take advantage of the $\theta$-independent sensitivity of the coherent state to steer, via a phase feedback, 
the squeezed state toward its optimal sensitivity point (namely toward the equatorial plane of the Bloch sphere in Fig.~\ref{fig1}).   
Proposals based on the interrogation of several atomic ensembles affected by the same LO noise have considered either classical~\cite{BorregaardPRL2013b, RosenbandARXIV2013, HumePRA2016}
or GHZ/NOON highly entangled states~\cite{KesslerPRL2014, HumePRA2016}.
Our hybrid coherent-squeezed protocol overcomes the SQL up to the Heisenberg scaling
$\Delta \theta \sim N^{-1}$
using robust states that are now routinely created in labs with large number of atoms, without 
requiring entanglement between different atomic ensembles or the implementation of non-destructive measurements.

In the following we illustrate in detail the protocol and compare analytical calculations of the Allan variance
obtained including realistic correlated $1/f$ noisy LO with 
numerical simulations of the full clock operations.
  
{\it Stability of a hybrid coherent-squeezed clock.}
A passive atomic clock operates by locking -- via a feedback loop -- the frequency of a laser to the energy transition of a two-level atom~\cite{RiheleBOOK}. 
In the hybrid protocol illustrated in Fig.~\ref{fig1} two atomic ensembles each having $N$ atoms are interrogated by the same LO:
one ensemble is prepared in a coherent spin state made by uncorrelated atoms 
while the other ensemble is in a squeezed state. 
A Ramsey interferometric protocol is identically applied to each clock. 
A free precession about the $z$ axis of the Bloch sphere rotates the atomic states by the same angle 
$\theta = \int_T dt~\delta \omega_{\rm LO}(t)$ during the time $T$, where $\delta \omega_{\rm LO}(t) = \omega_{\rm LO}(t)- \omega_0$, 
$\omega_{\rm LO}(t)$ is the time-dependent locked LO frequency and $\omega_0$ is the reference atomic frequency. 
After phase accumulation (which is the same in both ensembles), both states are rotated around the $x$ axis by an angle $\pi/2$.

The central operation of the hybrid protocol is the phase feedback, see Fig.~\ref{fig1}. 
First, the relative population difference $\J_z^{(1)}$ between the two clock levels in the first ensemble is measured \cite{Nota1}. 
Given the result $-N/2\leq \mu^{(1)} \leq N/2$, the phase is estimated as 
$\theta_{\rm est}^{(1)} = \arcsin(\mu^{(1)}/\mean{\J_x^{(1)}})$.
The phase feedback is implemented by a rotation of the second ensemble by the angle $\theta_{\rm est}^{(1)}$ about the $y$ axis.
We assume that measurement, estimation and feedback rotation all require 
a negligible time to be completed (we will release this assumption below). 

The key point of this proposal is that the phase-feedback rotates the squeezed state 
toward the equatorial plane of the Bloch sphere within a region of the order of $1/\sqrt{N}$ about the optimal phase estimation value
$\theta_{\rm opt}=0$, see Fig.~\ref{fig1}.
The measurement of the relative number of particles in the second ensemble, with results $\mu^{(2)}$\new{,} provides an 
estimate $\theta_{\rm est}^{(2)}= \arcsin(\mu^{(2)}/\mean{\J_x^{(2)}})$ of the phase $\theta - \theta_{\rm est}^{(1)}$.
We emphasize here that $\theta_{\rm est}^{(1)}$ plays the role of a stochastic but exactly 
determined number that has to be added to the unknown stochastic value $\theta$ which we want to estimate.
In other words, the estimation uncertainty of $\theta_{\rm est}^{(1)}$ is irrelevant in our protocol and 
it does not propagate to the error of the final estimation. 
Therefore, we are interested in $\theta_{\rm est} =  \theta_{\rm est}^{(1)} + \theta_{\rm est}^{(2)}$ which estimates
$\theta$ with a sensitivity given by $(\Delta \theta_{\rm est})^2  = (\Delta \theta_{\rm est}^{(2)})^2$.
At the end of the joint interrogation the LO frequency is steered by
$\omega_{\rm est} = \theta_{\rm est}/T$. This final step completes a single Ramsey cycle. 

The stability of the protocol is quantified by the Allan variance~\cite{RiheleBOOK}
$\sigma^2 = \tfrac{1}{2} {\mathcal{E}}[(y_2 - y_1)^2]$, with 
$y_2 - y_1 = \tfrac{1}{N_c} \sum_{n=1}^{N_c} (y_{n+1} - y_{n})$,
where ${\mathcal{E}}$ indicates statistical averaging and $N_c$ is the number of cycles (done in a total time $\tau = N_c T$). 
The quantity $y_{n} =  \frac{\Delta \theta_{\rm est}(T_n)}{\omega_0 T}$, with 
$\Delta \theta_{\rm est}(T_n) = \theta(T_n) -\theta_{\rm est}(T_n)$,
is the difference between 
the phase shift at the $n$th Ramsey cycle
$\theta(T_n) = \int_{T_{n-1}}^{T_n} dt~\delta\omega_{\rm LO}(t)$
and its estimated value $\theta_{\rm est}(T_n)$.
In the following we study the Allan variance in presence of LO fluctuations. 
We consider the unlocked LO $\tilde{\omega}_{\rm LO}(t)$ with power spectrum $S(f)=1/f$~\cite{RiheleBOOK}, where
$\delta \tilde{\omega}_{\rm LO}(t) = \tilde{\omega}_{\rm LO}(t) - \omega_0$ has a
time-independent variance ${\mathcal{E}}[\delta\tilde{\omega}_{\rm LO}(t)^2] = \gamma_{\rm LO}^2$ and zero mean $\mathcal{E}[\delta \tilde{\omega}_{\rm LO}(t)] = 0$.
These fluctuations are a most important source of noise
in realistic clocks, where atomic decoherence typically occurs on longer time scales. 
We now derive an approximated analytical expression for the Allan variance of the noisy clock that will be compared to full numerical simulations.  

We have verified numerically that statistically-averaged time correlations of the LO noise can be neglected (see also~\cite{AndrePRL2004,BorregaardPRL2013, KesslerPRL2014}),
${\mathcal{E}}[y_n y_m] \simeq  {\mathcal{E}}[y_{n}^2] \delta_{n,m}$, where $\delta_{n,m}$ is the Dirac delta function,
thus obtaining
\be \label{sigmay}
\sigma^2 = \frac{\mathcal{E}[(\Delta \theta_{\rm est})^2]}{\omega_0^2 \tau T}.
\ee
We recall here that $(\Delta \theta_{\rm est})^2  = (\Delta \theta_{\rm est}^{(2)})^2$ and the 
statistical averaging can be evaluated as $\mathcal{E}[(\Delta \theta_{\rm est}^{(2)})^2]= \int d \phi P(\theta_2) (\Delta \theta_{\rm est}^{(2)})^2$, 
where $(\Delta \theta_{\rm est}^{(2)})^2 = \sum_{\mu} (\theta_{\rm est}(\mu^{(2)}) - \theta_2)^2 P(\mu^{(2)} \vert \theta_2)$ is the estimator variance, 
$P(\mu^{(2)} \vert \theta_2)$ is the conditional probability to obtain the measurement result $\mu_2$ for 
a given stochastic $\theta_2 = \theta - \theta_{\rm est}^{(1)}$ with distribution $P(\theta_2)$.
In the following we calculate Eq.~(\ref{sigmay}) taking $P(\theta_2)= e^{-\theta_2^2/(2\kappa^2)}$, where the width $\kappa$
quantifies the quantum phase noise.
From error propagation we have  
\be \label{Dtheta}
(\Delta \theta_{\rm est}^{(2)})^2 = \frac{\Delta^2 \J_y}{\mean{\J_x}^2} +  \frac{\Delta^2 \J_x}{\mean{\J_x}^2} \theta_2^2.
\ee
The spin moments on the r.h.s of Eq.~(\ref{Dtheta}) can be calculated analytically with 
$\ket{\psi} \sim \int d \mu ~e^{-\mu^2/(s^2N)} \ket{\mu}_y$ as input state, where 
$s$ is a squeezing parameter ($s<1$ for spin-squeezed states) and
$\ket{\mu}_y$ are eigenstates of $\J_y$.
We get $(\Delta \J_y)^2 = s^2 N/4$,
$\mean{\J_x} = \frac{N}{2} e^{-1/(2 s^2 N)}$ and 
$\mean{\J_x^2} =  \frac{N^2}{8} ( 1+ e^{-2/(s^2N)} )$~\cite{Nota2}.
Replacing into Eqs.~(\ref{Dtheta}) and (\ref{sigmay}),
for $s^2N\gg1$~\cite{Nota3} we obtain 
\be \label{Allanpred}
\sigma^2 =  
\frac{1}{\omega_0^2 \tau T} \bigg( \frac{s^2}{N} + \frac{\kappa^2}{2 s^4 N^2} \bigg),
\ee
which manifests the interplay between squeezing $s$ and phase noise $\kappa$ in the Allan variance.
The optimization of Eq.~(\ref{Allanpred}) over the squeezing parameter $s$ 
gives
\begin{align}
& s_{\rm opt} = (\kappa^2/N)^{1/6}, \label{optimals}\\
& \sigma_{\rm opt}^2 = \frac{3}{2} \frac{1}{\omega_0^2 \tau T}  \bigg( \frac{\kappa^2}{N^4} \bigg)^{1/3}. \label{optimalAll} 
\end{align}
In absence of fringe hops, namely when $\theta(T_n) \lesssim \pi/2$, 
the fluctuations of $\theta_2 = \theta- \theta_{\rm est}^{(1)}$ are dominated by the quantum phase 
noise 
$\kappa^2=1/N$.
Replacing this value into Eq.~(\ref{optimalAll}) gives
\be \label{Allanhyb}
\sigma_{\rm opt}^2 = \frac{3}{2} \frac{1}{\omega_0^2 \tau T} \frac{1}{N^{5/3}}.
\ee 
This is the most important result of this paper for the hybrid clock of Fig.(\ref{fig1}), which is based on the interrogation
of two atomic ensembles. This result will be extended to the case of a larger number of atomic ensembles in the next sections. 

Equation~(\ref{Allanhyb}) shows a scaling of the stability with respect to the number of atoms $N$ that is superior 
to the one reached by optimal spin-squeezed states in a single Ramsey clock,  
$\sigma_{\rm sq}^2 \sim N^{-4/3}$~\cite{AndrePRL2004}.
It is worth to briefly elaborate here on this point. 
For large squeezing the second term in r.h.s. of Eq.~(\ref{Allanpred}) becomes significant 
due to the bending of the state in the equatorial plane of the Bloch sphere and strongly depletes the 
sensitivity except in a restricted  phase interval around $\theta_{\rm opt} = 0$ where can we still expect sub SQL. 
High sensitive atomic clocks works in regimes of large Ramsey times, which means large (stochastic) phase shifts $\theta \lesssim \pi/2$. 
Therefore, a single Ramsey clock operating with a squeezed state would mostly explore sub-optimal phase sensitivity regions eventually 
providing only modest improvements with respect to the SQL~\cite{AndrePRL2004}. 
The sub-SQL region shrinks with $s$ such that only modest squeezing parameters
can be used for the single Ramsey clock.
In contrast, the hybrid clock protocol allows the use states having much stronger squeezing ($s \ll 1$):
Eq.~(\ref{optimals}) predicts $s_{\rm opt} = N^{-1/3}$~\cite{Nota5}. 

When $\theta(T_n) \gtrsim \pi/2$ 
fringe hops (or phase slips) occur, which bias the estimation. 
In this regime $y_n \approx \gamma_{\rm LO}^2 T^2$ (for $1/f$ noise) and we find 
\be \label{sigmay_hop}
\sigma^2 = \frac{\gamma_{\rm LO} T}{\omega_0^2 \tau}.
\ee
The crossover from Eq.~(\ref{sigmay}) to Eq.~(\ref{sigmay_hop}) is signalled by an upward bending of the Allan variance as a function of $T$.

\begin{figure} [t!]
\centering
\includegraphics[width=0.9\columnwidth]{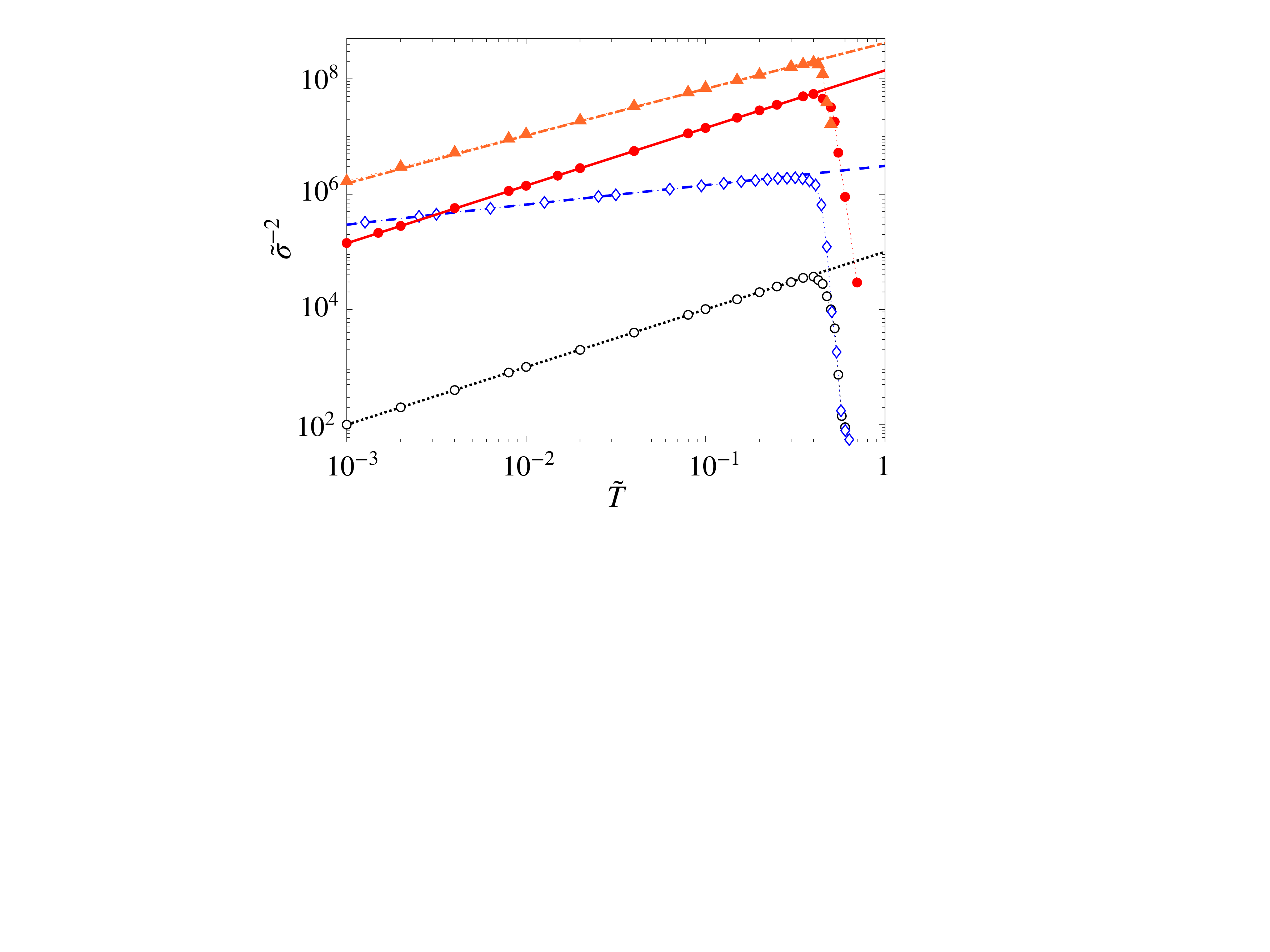}
\caption{Inverse Allan variance as a function of the interrogation time $T$ for the hybrid protocol implemented with {\it i})
a coherent state and an optimal squeezed state (red circles) with the solid red line being Eq.~(\ref{Allanhyb}) and 
{\it ii})  optimal spin-squeezed states (orange triangles) with the dot-dashed orange line being Eq.~(\ref{Allansqsq}). 
For comparison, we show the inverse Allan variance for a single Ramsey clock with a coherent spin state (black dots) and an optimal spin-squeezed states (green diamonds).
The black dot-dashed line is the SQL Eq.~(\ref{SQL}), while the dotted green, Eq.~(\ref{Allansingle}).
Here we have used rescaled units $\tilde{T} = \gamma_{\rm LO} T$, $\tilde{\sigma} = \sigma \omega_0 \sqrt{\tau/\gamma_{\rm LO}}$, and $N=10^5$.}
\label{fig2}
\end{figure}

We have numerically simulated the clock operations including the $1/f$ LO noise and averaging over several Ramsey cycles~\cite{Nota4}. 
In Fig.~\ref{fig2} we show the stability of the hybrid protocol as a function of the interrogation time $T$, for 
$N=10^5$ and an optimal squeezing parameter. 
Equation~(\ref{Allanhyb}), solid line, is in excellent agreement with the numerical results up to 
$\gamma_{\rm LO} T_{\rm opt} \approx 0.4$, where phase slips become likely and deplete the stability.
 The hybrid interrogation increases by order of magnitudes the long-term stability of a standard atomic clock operating with a coherent spin state on $N$ particles (black dots), 
 where the optimal interrogation time is also given by $\gamma_{\rm LO} T_{\rm opt} \approx 0.4$.
The dotted black line is the SQL Allan variance~\cite{ItanoPRA1993}
\be \label{SQL}
\sigma^2_{\rm SQL} = \frac{1}{\omega_0^2 \tau T N}
\ee
The hybrid scheme overcomes the long-term stability reached by the single atomic clock operating with spin-squeezed states optimized for each Ramsey time $T$ (blue diamonds), 
the blue dashed line being 
\be \label{Allansingle}
\sigma_{\rm sq}^2  = \frac{3}{2} \frac{1}{\omega_0^2 \tau} 
\bigg( \frac{\gamma_{\rm LO}^2}{TN^4} \bigg)^{1/3}, 
\ee
obtained for an optimal squeezing parameter $s_{\rm sq} = (\gamma_{\rm LO}^2 T^2 /N)^{1/6}$.
The hybrid protocol leads to a smaller Allan variance than the single Ramsey clock with optimal squeezing for an interrogation time $\gamma_{\rm LO} T \geq  1/\sqrt{N}$ that 
is obtained by comparing Eqs.~(\ref{Allansingle}) and (\ref{optimalAlln}) and is vanishingly small when increasing $N$.

{\it Extended hybrid protocol.}
We now extend the hybrid clock considering three or more atomic ensembles having the same number of atoms $N$ and interrogated by the same LO. 
In this case, the estimated phase from the second Ramsey clock, $\theta_{\rm est}^{(2)}$, is used to phase-feedback a third ensemble, and so on. 
In a cascade of $\nu$ squeezed-state ensembles all implemented with the same atom number $N$, 
the optimal value of the squeezing for the $\nu$th interferometer is
\be \label{optimalsn}
s_{\nu,{\rm opt}} = \bigg( \frac{3}{2}\bigg)^{\tfrac{1}{4}\big(1-\tfrac{1}{3^{\nu-1}}\big)} \frac{1}{N^{\tfrac{1}{2} \big(1-\tfrac{1}{3^\nu} \big)}},
\ee
which provides the optimized Allan variance
\be \label{optimalAlln}
\sigma_{\nu,{\rm opt}}^2 = \bigg( \frac{3}{2}\bigg)^{\tfrac{3}{2} \big(1-\tfrac{1}{3^{\nu}}\big)} \frac{1}{\omega_0^2 \tau T}  \frac{1}{N^{2-3^{-\nu}}}.
\ee
Equation (\ref{optimalAlln}) quickly approaches the Heisenberg scaling $\sigma_{\nu,{\rm opt}}^2 \sim 1/(TN^2)$, when $3^\nu\gg1$, with a prefactor $(3/2)^{3/2} \approx 1.84$. This is the second main result of this paper which extend Eq.~(\ref{Allanhyb}) to the case of more than two atomic ensembles. In the case $\nu=0$, we recover Eq.~(\ref{SQL}) for $s_{0,{\rm opt}}=1$, while in the case $\nu=1$ we recover Eq.~(\ref{Allanhyb}).
Notice that $s_{\nu,{\rm opt}}^2N \sim N^{1/3^\nu}$, which is consistent with the assumption $s^2N\gg1$ leading to Eq.~(\ref{Allanpred}) asymptotically in $N$. 
Increasing $\nu$, optimal states are more and more squeezed, Eq.~(\ref{optimalsn}), 
which corresponds to an increase of long-term stability compared to Eq.~(\ref{Allansingle}).

\begin{figure} [t!]
\centering
\includegraphics[width=1\columnwidth]{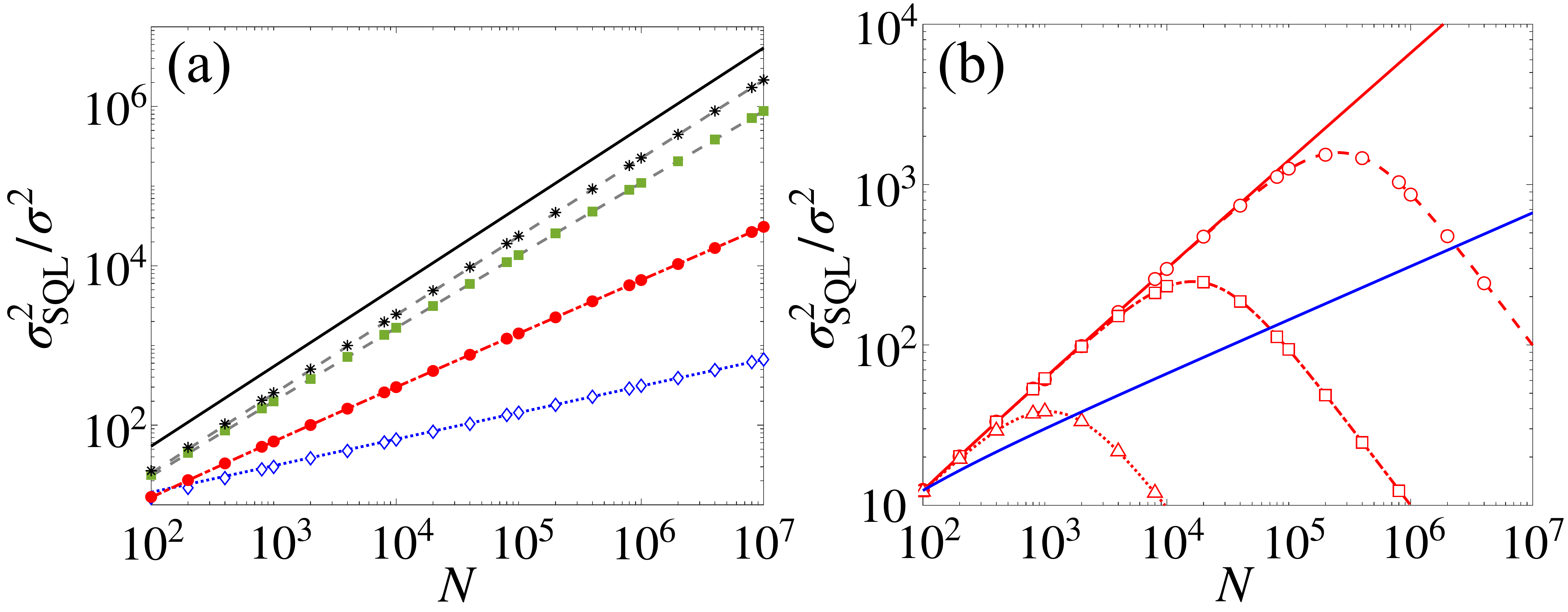}
\caption{(a) Inverse Allan variance as a function of the number of particles $N$.
Symbols are numerical results for the single Ramsey clock with optimal squeezed states (blue diamonds) and the hybrid coherent-squeezed protocol with 
$\nu=1$ (red dots), $\nu=2$ (green squares) and $\nu=3$ (black asterisks). 
Dashed grey lines are Eq.~(\ref{Allanpred}) including exponential corrections, see \cite{Nota3},
the dotted blue line is Eq.~(\ref{Allansingle}) for the single Ramsey clock, while the dot-dashed line is Eq.~(\ref{optimalAlln})
for the hybrid coherent-squeezed clock.
The thick black line is the asymptotic $\nu = \infty$ prediction of Eq.~(\ref{optimalAlln}).
(b) Inverse Allan variance of the hybrid coherent-squeezed protocol in the presence of a phase noise $\kappa^2_{\rm noise}$ in the implementation of the phase feedback:
$\kappa^2_{\rm noise} = 10^{-5}$ (triangles), $10^{-7}$ (squares) and $10^{-9}$ (circles). 
Lines are analytical predictions Eq.~(\ref{Allannoise}).
The solid blue line is Eq.~(\ref{Allansingle}) and the solid red line is Eq.~(\ref{Allanhyb}).
In both panels $\gamma_{\rm LO}T = 0.1$.}
\label{fig3}
\end{figure}

In Fig.~\ref{fig3}(a) we show that scaling of the inverse Allan variance as a function of the number of particles $N$, 
at $\gamma_{\rm LO} T=0.1$ and optimal values of the squeezing parameter (determined numerically).
For relatively small $N$, numerical results agree very well with an extension of Eq.~(\ref{Allanpred})
including exponential corrections, see \cite{Nota3} (grey lines), that agree with Eq.~(\ref{optimalAlln}) asymptotically in $N$.

{\it Stability of a hybrid squeezed-squeezed Ramsey clock.}
As shown in Fig.~\ref{fig2}, for very short interrogation times, 
the stability of the hybrid coherent-squeezed protocol is surpassed by the single Ramsey clock using optimal squeezed states.
We thus consider here a hybrid scheme where the first ensemble is in a squeezed state rather than a coherent state as considered above.
With a cascade of $\nu$ interferometers and optimizing the squeezing of all ensembles leads to
\begin{align}
&s_{\nu,{\rm opt}} =  \bigg( \frac{3}{2}\bigg)^{\tfrac{1}{4}\big(1-\tfrac{1}{3^\nu}\big)} 
\frac{(\gamma_{\rm LO} T)^{\tfrac{1}{3^{\nu+1}}} }{ N^{\tfrac{1}{2} -\tfrac{1}{3^{\nu+1}}}},\\
&\sigma_{\nu,{\rm opt}}^2 = \bigg( \frac{3}{2}\bigg)^{\tfrac{3}{2}\big(1-\tfrac{1}{3^{\nu+1}}\big)}  
\frac{1}{\omega_0^2 \tau T}
\frac{(\gamma_{\rm LO} T)^{2/3^{\nu+1}}}{N^{2-2/3^{\nu+1}}}.  \label{Allansqsq}
\end{align}
Equation~(\ref{Allansingle}) is recovered for $\nu=0$.
In Fig.~\ref{fig2} we show the numerical results for the stability of the hybrid squeezed-squeezed interferometer $\nu=1$ with optimal squeezed states (orange triangles), 
the orange dot-dashed line being Eq.~(\ref{Allansqsq}).
The hybrid scheme with optimal squeezed states always overcomes the stability of the single Ramsey interferometer with optimized squeezing.

{\it Impact of imperfections.}
The phase feedback operation is the key element of the hybrid protocol.
We take here into account imperfections of the phase feedback, either due to a dephasing of the 
LO during the time needed for the implementation of the feedback, or an imperfection in the rotation angle.    
Assuming a Gaussian dephasing with width $\kappa_{\rm noise}^2$, the Allan variance becomes
\be \label{Allannoise}
\sigma^2 = \sigma_{\rm opt}^2 + \kappa_{\rm noise}^2.
\ee
In Fig.~\ref{fig3}(b) we plot the numerical results for the Allan variance for the coherent-squeezed hybrid protocol in the presence of a finite dephasing.
The possible advantage of the hybrid protocol over the single Ramsey clock with optimal squeezed states relies on the condition $\kappa_{\rm noise}^2 \lesssim \sigma_{\rm opt}^2$.
To conclude, it is worth to notice that, since the readout of the squeezed state ensemble is close to the optimal phase point, we can make use on non-linear readout schemes 
to overcome detection noise~\cite{DavisPRL2016, FrowisPRL2016, MacriPRA2016, HostenSCIENCE2016}. 
It is also possible to combine the present hybrid scheme with different protocols proposed in the literature to 
increase the Ramsey interrogation time \cite{BorregaardPRL2013b, RosenbandARXIV2013, HumePRA2016, MullanPRA2014, Wu, LerouxMETROLOGIA2017}

{\it Conclusions.}
We have proposed a clock protocol based on the joint use of a coherent spin state and one (or more) squeezed spin state(s) interrogating the same noisy local oscillator. 
The interrogation of the atoms in the coherent spin state
allows to steer the spin squeezed states toward their 
optimal phase sensitivity point. 
Our clock consists on a hybrid combination of a classical and a quantum interferometric implementation.
The basic principle has some analogies with 
the hybrid quantum computation~\cite{McCleanNJP2018, MollQST2018} and simulation~\cite{PeruzzoNATCOMM2014, KokailNATURE2019}
protocols recently explored in the literature, where classical algorithms were combined with 
quantum resources to speed up specific tasks.  
 
The clock operations discussed in this manuscript, including the creation of atomic spin-squeezed states, 
are within the current technology.

\begin{acknowledgments}
{\it Acknowledgments.---}
We acknowledge funding of the project EMPIR-USOQS, EMPIR projects are co-funded by the European Union’s Horizon2020 research and 
innovation programme and the EMPIR Participating States.
We also acknowledge support by the H2020 QuantERA ERA-NET cofund QCLOCKS.
\end{acknowledgments}


\begin{thebibliography}{400}
\bibitem{Nota1}
Here, $\hat{\vect{J}} = (\J_x, \J_y, \J_z)$ is the collective spin of the atomic ensemble, 
where $\J_{\vect{n}} = \sum_{j=1}^N \hat{\sigma}^{(j)}_{\vect{n}}/2$ and $\hat{\sigma}^{(j)}_{\vect{n}}$ is a Pauli operator along a direction $\vect{n}$ for the $j$th qubit.
\bibitem{Nota2}
These equations are obtained considering $\mu$ as a continuous variable, 
which is accurate for $N \gg1$ and $s^2 N \gtrsim 1$, and up to corrections of the order $1/N$.
\bibitem{Nota3}
The equation $\sigma^2 =  \frac{2 s^2 + N ( 1+ e^{-2/(s^2 N)} -2 e^{-1/(s^2 N)} ) \kappa^2}{2N e^{-1/(s^2 N)}}$ is valid for $s^2 N \gtrsim 1$, while
Eq.~(\ref{Allanpred}) is obtained from a Taylor expansion of the exponentials and keeping the leading order in $1/(s^2 N)$.
\bibitem{Nota5}
Still higher squeezing can be exploited if the number of 
atoms in the spin-coherent atomic ensemble is higher than the number of atoms in the second ensemble.
\bibitem{Nota4} None of the approximations used to derive the analytical expressions is used in the numerics. 
We typically consider $N_c = 10^4$ Ramsey cycles and average the inverse Allan variance $1/\sigma^2$
over 20 independent realizations. 
Temporal correlations of the LO noise are included in the numerics.
\end{thebibliography}
\end{document}